\documentclass[12pt,preprint]{aastex}  
\begin{document} 

\title{{\ion{H}{1}} velocity dispersion in NGC 1058}
\author{A. O. Petric}
\affil{Astronomy Department, Columbia University, New York, NY USA \\
andreea@astro.columbia.edu}
\author{M. P. Rupen}
\affil{National Radio Astronomy Observatory, P.O. Box O, Socorro, NM,
87801, USA \\
mrupen@aoc.nrao.edu}
\begin{abstract}
We present excellent resolution and high sensitivity Very Large Array (VLA)
observations of the 21cm {\ion{H}{1}} line emission from the face-on galaxy
NGC 1058, providing the first reliable study of the {\ion{H}{1}} profile shapes
throughout the entire disk of an external galaxy. Our observations show an 
intriguing picture of the interstellar medium; throughout this galaxy velocity--
dispersions range between 4 to 15 km $\rm{s} ^{-1}$ but are not correlated
with star formation, stars or the gaseous spiral arms. The velocity 
 dispersions decrease with radius, but this global trend has a large
 scatter as there are several isolated, resolved regions of high dispersion.
 The decline of star light with radius is much steeper than that of the velocity
 dispersions or that of the energy in the gas motions.
\end{abstract}


\section{Introduction}
Observations of {\ion{H}{1}} velocities perpendicular to the disk ($v _{z}$) are 
necessary for studies of both the interstellar medium (ISM) (McKee \& Ostriker 1977, Kulkarni \& Heiles 1988, Braun 1992, 1997) 
and disk dynamics (Oort 1932; Rupen 1987; Lockman \& Gehman 1991; Merriefield 1993; Malhotra 1994, 1995; 
Olling 1995) because they set a direct upper limit on the thermal and kinetic temperature of the gas. 
Hence the {\ion{H}{1}} velocities
perpendicular to the disk are an important dynamical tracer and as such can be used to constrain, 
both the gas mass distribution in the plane and its vertical structure (i.e., density as a
 function of height-$z$ above the plane)(van der Kruit \& Shostak 1982,1984, Lockman \& Gehman 1991, Malhotra 1995). 

 The behavior of the velocity dispersions as a function of 
galactic radius is important for determinations of the shape of dark matter halos. To date
 even the most sophisticated methods (e.g. Olling 1995, 1996) assume either a constant or an 
azimuthally symmetric velocity dispersion. Our measurements can therefore be used with studies of
 edge-on systems to determine radial variations in the mass to light (M/L) ratio.

 Processes associated with star formation,such as stellar
winds and multiple supernova explosions are thought to put energy into the ISM in the form of mechanical
 energy, starlight (which leads to photoelectric emission from dust grains), and cosmic rays. The velocity dispersion in the 
$z$ direction is intimately connected to the forces holding the gas against gravitational instabilities 
and hence to star-formation in the disk (e.g. Mac Low \& Klessen 2004, Li, Mac Low \& Klessen 2005).
Measuring the degree of correlation between 
the locations of star-forming regions and those of high dispersion is a good method to investigate
the relation between star-related energy sources and {\ion{H}{1}} bulk motions.

The face-on spiral galaxy NGC 1058 (e.g. Eskridge et al. 2002) is ideal for studies of {\ion{H}{1}} 
  $v_z$ dispersions. Its low inclination (4\arcdeg--11\arcdeg) (Lewis 1987, van der Kruit \& Shostak 1984)
 means that the gradient in rotational velocity is small across the beam and therefore it does not significantly 
corrupt measurements of velocities perpendicular to the disk. 

Single dish studies of NGC 1058 (Allen \& Shostak 1979, Lewis 1975, Lewis 1984) lack the resolution
to trace the dispersion across the disk but through modeling of the rotational component these authors 
estimate it to range between 7 and 9 km/sec.
In a series of papers (1982-1984) van der Kruit \& Shostak analyze {\ion{H}{1}} emission profiles in a number of 
face-on galaxies and determine that the velocity dispersions in NGC 1058 range only between 7 to 8 km/sec at all radii
with very little variation. Dickey, Hanson, \& Helou (1990) find the that velocity dispersion in NGC 1058 decreases with optical surface brightness but that in the extended gas disk, beyond the Holmberg radius the velocity dispersion is {5.7 km/sec} everywhere, that is no variations with spiral phase or {\ion{H}{1}} surface density are found. 

 All previous determinations of the {\ion{H}{1}} velocity dispersion in NGC 1058 have
 been hindered by low spatial (e.g., Lewis 1984) and/or
        spectral (e.g., van der Kruit \& Shostak 1984) resolution
        as well as by relatively poor sensitivity, requiring
        smoothing over large sections of the galactic disk, or missing
        up to 40$\% $ of the total flux (Dickey, Hanson, \&  Helou
        1990). These trade-offs have led to significantly
        different conclusions about the {\ion{H}{1}} velocity dispersion.
 Our sensitive observations at high spatial and velocity resolution as 
 well as recovery of the entire single dish flux, allowed us to accurately
 measure the profile widths even in the outskirts of the {\ion{H}{1}} disk,
 to resolve the arm from the interarm regions, and to analyze in detail the 
{\ion{H}{1}} profile shapes, not just their breadths throughout the {\ion{H}{1}} disk. 

\section{Observations and Data Reductions}
The 21cm line of neutral hydrogen in NGC 1058 was observed
        with the VLA in the C and  CS\footnote{The CS (shortened C)
configuration moves two antennas from intermediate
 stations in the standard C configuration to the center of the array. The resulting short 
spacings significantly increase the sensitivity of the array to extended structure,
 while maintaining the same spatial resolution (Rupen 1997). }configurations.
The C configuration data was taken on 14 and 15 June 1993 for a total time on-source of 12.23
hours. The D configuration observation were performed on 7 and 8 November 1993 for a total time 
on source of 2.67 hours, and CS configuration data was collected on January 3, 1995 for a
total time on-source of 5.42 hours. Rupen (1997,1998) gives a detailed account of the UV coverage 
in each of the configurations, compares the merits of each configuration, and discusses the benefits
of combining them. 

Both the C and CS configurations have a
 maximum baseline of 3.6~km, while the D configuration has a maximum baseline 1~km). The minimum baseline,
 which determines the size of the most extended feature
which can be observed by the VLA, is 35~m. 
 All observations were taken in dual polarization mode and Hanning
smoothing was applied on-line; resulting in
 127 independent spectral channels with a velocity width of 2.58 km/sec.
 We followed the normal AIPS calibration
procedures and used the same flux (3C48) and phase
(0234+285) calibrators throughout. The continuum emission was approximated as a 
linear fit to visibilities in 20 line-free channels on each side of the signal, and
this fit was then subtracted from the {\it {uv}}-data in all the channels. 
Rupen (1999), gives a detailed description 
of the bandpass calibration and continuum subtraction.  
             
        The data cube presented here was deconvolved using the CLEAN algorithm as implemented 
by AIPS task IMAGR.{\footnote{A description of IMAGR can be found in the AIPS cookbook 
available online at $http://www.aoc.nrao.edu/aips/cook.html$.}}, iterated until the residuals
 were nearly zero and the flux density in the CLEAN model was
stable. The cube was tapered to a resolution of  $30\arcsec \times 29\arcsec $, or
 $1.3\,\times\,1.3$ kpc at a distance of 
10 Mpc (Ferguson et al. 1998). Rupen (1997), presents a detailed comparison of several 
cleaning algorithms and motivates the use of the CLEAN algorithm for this data.
A more general discussion of CLEAN as implemented in AIPS is given in chapter 5, of the AIPS
cookbook as well as in Cornwell, Braun, \& Briggs 1999. A more specific examination of
 deconvolution algorithms as applied on our NGC 1058 data is presented in  Rupen (1997,1999). 

  The RMS noise level in the line channels 
of the cube was 0.5 and mJy/beam, corresponding to a column density of $1.6 \times 10^{18} $
cm$^{-2}$ per channel. The {\ion{H}{1}} integrated line profile agrees with those obtained in
single dish studies (Allen $\& $ Shostak 1979) after both the single dish and the VLA data are
 corrected for primary beam response.  

Figure 1 presents the frames of the 30\arcsec ~data cube. Each image (traditionally called a channel
map) in this figure represents the 21 cm line emission at a certain velocity, the abscisa and ordinate
axis are the RA and Dec coordinates. The lack of artifacts in these images (such as a negative bowl around
the galaxy) also suggests that the images have been correctly deconvolved. 

\section{Results and Analysis}
The general properties of NGC 1058 are presented in Table 1. 
 Figure~\ref{fig:M1contM0grey} shows intensity weighted mean velocity
contours atop the {\ion{H}{1}} intensity map and presents
 the {\ion{H}{1}} spiral structure of NGC 1058. Figure~\ref{fig:M1contM0grey} 
also illustrates the superb sensitivity and resolution of these studies,
which allowed us to measure the {\ion{H}{1}} emission at distances of 
approximatively 10~kpc from the center of the disk and to differentiate between the arms and the inter-arms. We characterize the widths
of the {\ion{H}{1}} profiles by the relative dispersions $\sigma _{\rm{v}}$
of the best Gaussian fit \footnote{ The FWHM is often use to characterize
  the \ion{H}{1} line widths. This {\it{tradition}} is based upon the
  fact that \ion{H}{1} profiles are modeled by one or multiple
  Gaussians, where the flux as a function of velocity $v$ is given by $$f(v)={1\over{\sigma\sqrt{2\pi}}}exp\left
    [-\left(1\over{2{\sigma _\mathrm{v}}^2}\right){(v-v_0)}^2\right
  ]$$ and $v_0$ is the velocity at associated with the peak flux.}; the fits
were done using a least squares minimization algorithm.

 Figures ~\ref{fig:prof45_1058} and ~\ref{fig:prof30_1058}, show the observed profiles
        for a few pixels throughout NGC~1058 from the 45\arcsec ~and
        the 30\arcsec ~data sets respectively. The single Gaussians
        which best approximate the shapes of these profiles, as well as their residuals are also
        shown. Note, that each of the profiles is representative of
        the remainder of the spectra, it is not a {\it{best find} } or 
        the result of averaging over large areas of the disk or
        velocity space.

 While the residual patterns suggest that a
        single Gaussian is not a good functional description of the
        \ion{H}{1} profiles in NGC 1058, the FWHM and $\sigma_v$
        derived from the single Gaussian fits track well the
        intrinsic width of the \ion{H}{1} spectrum. This proportionality allows
        us to describe the widths of the profile in terms of the results of our
        least squares fitting. The general characteristics of the velocity
        dispersion will be discussed in terms of the 45\arcsec and 30\arcsec ~cubes.
       
\section{General Characteristics of the Velocity Dispersion }    
   Figure~\ref{fig:SigCont1058c30} presents the distribution of velocity dispersions
across the disk of NGC 1058. Unlike previous observers of NGC 1058, we find a wide range of dispersions from
$~4$ to 14 km sec$^{-1}$ in addition to a few extremely narrow profiles with $\sigma _v ~\sim ~3.5$~
km sec$^{-1}$. These narraow profiles are found in regions of relatively low column density at 
radii greater than $300\arcsec $ or 13~kpc. 
There are three regions of high dispersion which stand out in Figure~\ref{fig:SigCont1058c30}: 
 one in the center (labeled C and $\sim 4.5$ kpc across) and two others symmetric about the center
 in the North-West (N $\sim 3$ kpc) and South-East (S $\sim ~3\times 5$ kpc) of the center.
We find no obvious correlation between high {\ion{H}{1}} velocity dispersion and
stars or star formation tracers such as $H_{\alpha}$ (Figure~\ref{fig:Halpha_on_Sig}), radio contiuum, SNe except in the central region C. 
The most probable explanation for the observed highest dispersions outside the central region
(i.e. in N and S) are small scale ($\leq 0.7$~kpc) bulk motions (see section 7 below). 
In the southern, part the disk could be warping
(van der Kruit \& Shostak 1984, Shen \& Sellwood 2006) leading to the observed broad profiles. 
Figure~\ref{fig:asym} shows that {\ion{H}{1}} profiles from N and S are also assymmetric. However, a 
similar explanation for region N would suggest rather impressive small-scale structure in 
the warp as it would requiere the inclination to change $\sim$~3 degrees over a region smaller
than 0.7 kpc in diameter, if we assume from Tully Fisher
an intrinsic rotation velocity of 150 km s$^{-1}$~. A more exciting alternative explanation for
the bulk motions observed in N is that they are caused by the infall of gas left over from galaxy formation. 
However, this is somewhat difficult to reconcile with the relatively low column density in these regions. 

Two global trends are evident from the derived dispersions: a radial fall-off,
shown in Figure~\ref{fig:SigvsR1058c30}, and a predominance of the broadest profiles 
in the inter-arm regions of the galaxy (Figure~\ref{fig:SigonM0c30}).
Ferguson et al. (1998) used deep $H\alpha$ observations to reveal
 the presence of \ion{H}{2} regions in the central 6 kpc of NGC
 1058.  There are several knots of high dispersion (12 to 13.5 km/sec) in
 region C, with most of the profiles measuring between 7.5 and 11
 km/sec. However, none of the star formation sites outside the central
 $\sim$2~kpc discovered  in that study seem to affect the width of
 the profiles. Also, regions N and S are located in the inter-arm regions and are not associated 
with sufficiently strong star formation to be detected in the Ferguson et al. (1998) study. 
Therefore we find that the dispersions do not correlate with star formation as shown from the overlay of Feruson et al (1998)'s H$\alpha$ map atop contours of velocity dispersion Figure~\ref{fig:Halpha_on_Sig}.

Figure~\ref{fig:ecomp}  shows the kinetic energy in the gas associated with motions perpendicular
to the disk. Because only a qualitative behaviour was of interest here,
 the kinetic energy in vertical motions at a certain pixel location was roughly
 approximated as the product of total intensity times the square of the velocity dispersion. 
Approximated as such, the kinetic energy in vertical motions does not follow
the decline in star light which drops with radius as $\sim \rm{exp}^{-{{r}\over{30\arcsec}}}$. 
Figure~\ref{fig:SigonM0c30} shows that, the broadest profiles seem to be found in relatively low 
        column density areas between the spiral arms (as traced by \ion{H}{1}). As such we do not find 
a correlation between the velocity dispersion of stars in the disk or the {\ion{H}{1}} column density.

The dissimilarity
between stars, star-formation, {\ion{H}{1}} intensity, and the kinetic energy in the gas
implies that processes other than those directly associated with stars put energy into the ISM. 
Sellwood $\&$ Balbus (1999) suggested that magnetic fields with strengths of a few micro-gauss
in these extended disks allow energy to be 
extracted from galactic differential rotation through MHD-driven turbulence. While that 
mechanism predicted a uniform dispersion outside of the optical disk, in an attempt to explain
 lower quality data on NGC 1058, a similar mechanism has the potential of explaining the level and 
behaviour of the velocity dispersions as a function of radius (Sellwood, private communication). The
Sellwood \& Balbus (1999) paper 
generated significant work on  numerical models that predict the occurance of the magnetorotational
instability in galactic disks (e.g. Dziourkevitch, Elstner, \& Rudiger 2004, Pionteck \& Ostriker 2004).

\section {Profile Shapes}
Any model that would explain how energy is put into the ISM must account for the shape of
the profiles in NGC 1058. A single Gaussian least-squares fitting routine was run on 
the 30\arcsec ~and 45\arcsec ~data sets. In both cases, we found that while the signal to noise for most 
profiles was excellent the chisq per degree of freedom was larger than a few,
the residuals also suggested that the wings were broader than those of a Gaussian.

To understand whether how the {\ion{H}{1}} line shapes varied throughout the galaxy,
the profiles were normalized by flux, aligned so that their peaks were at the 
same central velocity. These were plotted in units of FWHM (2.354$\sigma _{v}$),
 using the parameters from the single Gaussian fits to control the 
scaling. This was done to reveal only the difference in the line shapes and not other differences
such as the width of peak intensity. While stacking up the 45\arcsec ~profiles it became clear
 that almost all profiles appeared to have the same shape. Figure~\ref{fig:UPfid45} suggests that despite it being
non-Gaussian, the shapes of the line profiles are identical throughout most of the galaxy when 
scaled by $\sigma _v$ and their peak flux and aligned so that their peaks occur at the same 
velocity.

For the 45\arcsec ~data, median shapes from profiles within different width and peak
intensity ranges were compared and found to be identical within the error-bars. The method
used to derive such median line shapes is fairly straightforward. 
After the pixel selection (by FWHM, location in the galaxy,
etc.) the profiles corresponding to every pixel were normalized in
intensity dividing by the peak flux. A grid of 63 channels for the
45\arcsec \,\,data and 108 for the 30\arcsec \,\, data was set up to replace the
velocity axis from units of km/sec to units of FWHM. For example
suppose that the velocity corresponding to the peak of a certain
profile (in the 45\arcsec ~cube) is $V_{cen}$\,km\,s$^{-1}$ and that
its FWHM is {$FW$\,km s$^{-1}$}. Only the channels between
{$V_{cen}\,-\,3\times FW$} and {$V_{cen}\,+\,3\times FW$} were used in
deriving the median shape. The normalized
fluxes corresponding to these channels were then
resampled onto a grid where each bin (i.e.; channel) is 6\,$FW$ divided by the 
number of channels. Each bin therefore contains a certain number
distribution of normalized fluxes; these fluxes were then sorted and
the middle value is taken as the median.
 
Median profiles were also derived and compared from various areas throughout 
the disk and the line shapes appeared similar everyhwere except in N and S, where
the profiles where more asymmetric as previously discussed. For brevity we present just
two of these tests in Figure~\ref{fig:up45test}. The same 
median comparison tests were done on the 30\arcsec ~data. At the 30\arcsec ~resolution 
median profiles derived for certain ranges of peak flux
 and from various areas throughout the galaxy were also identical. However, 
the  median  profiles derived for various ranges of FWHM appeared to vary in the shape of their wings,
perhaps because of the lower signal to noise in this data set, and to the smaller number of
broad ($\sigma _v \geq 10 km ~s^{-1}$) than that of narrow lines.  

Throughout our analysis we assumed that the noise characteristic in each profile is
random and that the rms noise is the same regardless of the strengh of the
signal. This assumption need not be true as deconvolution algorithms
seem to produce noise that is proportional in a non-linear fashion
with signal (Rupen 97). However different noise for different flux levels will hardly 
lead to a universal, non-Gaussian line shape. A double Gaussian (a narrow and a broad component)
 as shown in Figure~\ref{fig:UP2G} is
a good fit to the median profile derived from the 45\arcsec ~data. 

\section{Kinetic Energy Distribution}
The uniformity of the profile shape in NGC 1058 suggests that on scales
 of 2.5~kpc, the neutral gas is
being stirred into the same distribution of energy per unit mass and that
this distribution is different than that for other galaxies (e.g. the
Milky Way). Figure~\ref{fig:UPenergy} shows the normalized kinetic
energy (KE) distribution for the
Milky Way and for NGC 1058. This comparison is only qualitative. The term ``normalized''
 in the case of the Milky Way refers to the fact that the KE distribution was
obtained from a model (i.e. double Gaussian fit) of the \ion{H}{1}
emission at the North Galactic Pole; this model was presented in
Kulkarni \& Fich (1985), hereafter KF85, and it only includes ``normal'' emission, i.e.
it does not include emission from the
\ion{H}{1} falling into the disk. These authors
 corrected for the infalling emission by assuming that the huge bump on one side of the profile 
represented infalling gas. To remove the bump they reflected the
profile about the velocity corresponding to the peak flux and obtained 
the profile shown in Figure~\ref{fig:UPenergy}. The units of the KF
plot are Kelvin$^2$ km$^2$ s$^{-2}$. The units for the NGC~1058 are
arbitrary, and the term ``normalized'' in this case means that instead of
flux or temperature, we use flux divided by peak flux, and bin numbers instead of 
velocities. To compare those qualitatively we
aligned the KF profile with the NGC~1058 median
profile from the entire 45\arcsec ~data set. The aligning
  was done by fitting a single Gaussian to the KF85 profile and to the
  NGC 1058 median profile. We require and that the limits of the KF85 and the NGC 1058 profiles
  span an equal number of FWHM.

The striking feature in the Galactic energy distribution, also noted 
by Kulkarni \& Fich (1985) is the almost constant kinetic energy for about
50\,km/sec. In contrast, NGC 1058's KE curve is more centrally
peaked.  Presumably the KE distribution is set both by the galactic
potential as well as explosive ISM events (such as SNe, star
formation, and infalling gas). It is not perfectly clear how these factors have
shaped the energy distribution as a function of velocity of either the Milky Way or NGC 1058.
function of velocity. 

\section{ Beam Smearing and Bulk Motions}
	An accurate study of the profile shapes throughout the galaxy require us to understand the effect of beam smearing on our measurements. Consider a round spiral disk with gas moving in circular orbits at a velocity $v_{circ}$; attaching polar coordinates to this disk ($r_d , \theta$) and letting the angle 
between the normal to the plane of the galaxy and line-of sight be referred to as the 
inclination angle ($i$) the observed radial velocity on a set of sky-coordinates ($x,y$) will
be
$$v_z(x,y)~=~v_z(r_d,\theta)cos(i)~+~v_{circ}(r_d)~sin(i)cos(\theta) + v_{red}$$ where $v_{red}$ is 
the velocity of the galaxy for which with respect to the observer. Obviouosly a smaller $i$ is
(a more face-on) makes it easier to measure the true $v_{z}$ distribution. A
gradient in the $v_{circ}(r_d)sin(i)cos(\theta)$ across the resolution element (beam) will 
increase the width of the profile and confuse the measurements of the velocities perpendicular to the disk.
This problem is known as beam smearing.

	Two tests were performed using our highest resolution (15\arcsec) data to quantify the 
effect of beam smearing on our measurements of $\sigma _v$. First, we determined the maximum 
in-plane velocity difference within a beam which would contribute to the width of the line
profile at a certain position in the galaxy (i.e. at a pixel). This was done by finding 
the maximum difference (hereafter $\rm{M_{diff}}$) between the central velocity $\rm{v_{cen}}$
 (i.e. the velocity associated with the peak flux as derived from the single
 Gaussian fit to the 15\arcsec ~data) of the {\ion{H}{1}} pixel and  the central velocities
 of all the pixels within a square with 32\arcsec ~sides centered on that pixel. 
Figure~\ref{fig:beam_smear1} shows the map of these maximum difference. 
This method is based on the assumption that differences in $\rm{v_{cen}}$ are due to gas motions
in the plane of the galaxy. This test shows that $\sigma _v$ is correlated with $\rm{M_{diff}}$.
The correlation between $\sigma _v$ and $\rm{M_{diff}}$ 
suggests the exitence of bulk motions on scales smaller and equal to those probed by our highest resolution data 15\arcsec ~(0.7 kpc). 

Finding $\rm{M_{diff}}$ across NGC 1058's disk gives an upper limit to 
the broadening of the {\ion{H}{1}} profiles. To better understand the effect
 of beam smearing on our observations we constructed a simple model of how {\ion{H}{1}} in NGC 1058 
would appear if it was an infinitely
cold disk; we then convolved this model with a 30\arcsec ~beam and ran our Gaussian 
least squares fitting routine on the resulting {\ion{H}{1}} profiles. The widths of these
final model profiles were significantly smaller than those measured in NGC 1058
 (Figure~\ref{fig:beam_smear2}) suggesting that beam smearing does not have a significant impact on our observations. 

\section{Summary and Conclusions}
         Excellent resolution and high sensitivity \ion{H}{1} observations of NGC 1058 show
         an intriguing picture of the interstellar medium
         throughout this galaxy: the velocity dispersion ranges 
         from 4 to 14 km/sec but is not correlated with star
         formation or the spiral arms, which is another major ISM regulator. 
         Global trends such as a radial fall-off must be explained in the
         context of significant local effects; most notable among
         these are isolated, resolved regions of high velocity dispersions as
         well as significant scatter in the dispersion at a given
         radius. In summary unlike some previous studies, we find
         that the dispersion is not constant and it does not simply
         decline with radius. we also find that there is no tight correlation between the width of
         the profiles and the spiral arms. 

         The most probable source for the highest dispersions observed 
         outside the central regions are small scale ($\leq$0.7 kpc) bulk
         motions. The energy sources supporting such motions are not
         entirely clear. The disk is warped in the southern part (van der
         Kruit \& Shostak 1984, Shen \& Sellwood 2006),
         leading to the observed broad profiles: however, a similar
         explanation for region N would suggest a rather impressive
         small-scale structure in the warp, as it would require the
         inclination to change by $\sim $ 3 degrees over a size
         smaller than 0.7~kpc.
         
         There is no obvious correlation with stars or star
         formation tracers such as H$\alpha$, radio continuum, SNe except in region 
         C; nor is it clear what role, if any, is played by spiral 
         arms in driving the observed small scale bulk motions. Some of the measured
 velocity dispersions are higher than the {10 km~$s^{-1}$}
         canonical sound speed in the ISM, but since we
         cannot easily measure directly the pressure and
         {3-dimensional} density structure of the gas, we cannot determine the
         exact sound speed to know if we are indeed seeing supersonic
         motions.
	
	The shapes of the {\ion{H}{1}} profiles in NGC 1058 are non-Gaussian and
hence cannot be explained as emission from single temperature gas. Therefore, it is not
clear whether these narrow profiles are evidence of a lower thermal balance point between
heating and cooling mechanisms in NGC 1058's outskirts as compared to the rest of the galaxy. 
    
A double Gaussian description of the \ion{H}{1} profile is far from a
complete surprise. The surprise is the constancy between the broad 
and narrow components throughout NGC~1058's \ion{H}{1} disk. In previous
studies (e.g. Mebold 1972, Young \& Lo 1996) it was found that some of the
\ion{H}{1} profiles were well described by double Gaussians, and these
associated the narrow Gaussians with the CNM and the broad with the
WNM. Young \& Lo (1996) found that the narrow component existed only in
regions of high \ion{H}{1} column density, next to areas with active
star formation. It is unlikely that the universal profile in NGC 1058
can be explained as a combination of cold and warm medium  for the
narrow and broad component respectively, because it seems difficult to
have the same ratio of warm to cold gas in regions associated with
stars and star fromation and at radii three times the optical
$R_{25}$. Also, high resolution observations in
other galaxies (Braun 1998) showed that the CNM 
dissappears at the edges of the optical disk. 

Further quests on the observational front such as (at what resolution will this universality break down, is
this spatial scale particular to NGC 1058, can we see the same shape and/or its
universality in other systems), as well as theoretical efforts to
model mechanisms of injecting energy into the ISM, and
 determine how that energy dissipates throughout a fractal
ISM are necessary to understand the full significance of the universal profile in our
45\arcsec ~data cube. 

A.P. would like to thank Jacqueline van Gorkom for invaluable help in
designing the experiment, as well as during during the analysis process
 and in editing this document. A.P. would also like to thank
Liese van Zee, Mordecai Mac-Low and Jennifer Donovan for their helpfull suggestions and discussions. 
The National Radio Astronomy Observatory is a facility of the National Science Foundation 
operated under cooperative agreement by Associated Universities, Inc..

\begin{deluxetable}{lc}
\tablecolumns{2}
\tablecaption{General Properties\label{tab:Gen_Prop_Gal}}
\tablehead{ \colhead{} & \colhead{NGC 1058}}
\startdata
R.A.(B1950)                            & 02 40 23.2\\
Dec (B1950)                            &+37 07 48.0\\
Morphological type                                       &Sc \\
V$_\mathrm{sys}$\,[km/sec]                                  & 518\\
L$_B$\,[L$_{\odot}$]        &$1.5 \times 10^9$\\
M$_{\sc{HI}}$[M$_{\odot}$]  &$2.3\times 10^9$\\
SFR\,[M$_{\odot}$\,yr$^{-1}$]\tablenotemark{a}      &$3.5\times 10^{-2}$ \\
D$_{25}\times$d$_{25}$\,[arcmin]\tablenotemark{b} &$3.0\times 2.8$\\ 
Distance[Mpc]\tablenotemark{c}                &10\\
Physical equivalent of 1\arcsec &48.5\,pc \\
Inclination\tablenotemark{d}                  &4--11$^{\circ}$\\
Environment\tablenotemark{e} &member of the NGC 1023 Group\\
\enddata
\tablenotetext{a}{SFR stands for Star Formation Rate, it was calculated
  from H$\alpha$ fluxes by Ferguson, Gallagher \& Wyse (1998)}
\tablenotetext{b}{NASA/IPAC extragalactic database (NED)}
\tablenotetext{c}{Ferguson, Gallager, \& Wyse (1998) }
\tablenotetext{d}{van der Kruit \& Shostak (1984) } 
\tablenotetext{e}{~Lewis (1975)}
\end{deluxetable}
\clearpage
\begin{figure}
\resizebox{\textwidth}{!}{
\rotatebox{-90}{
  \plotone{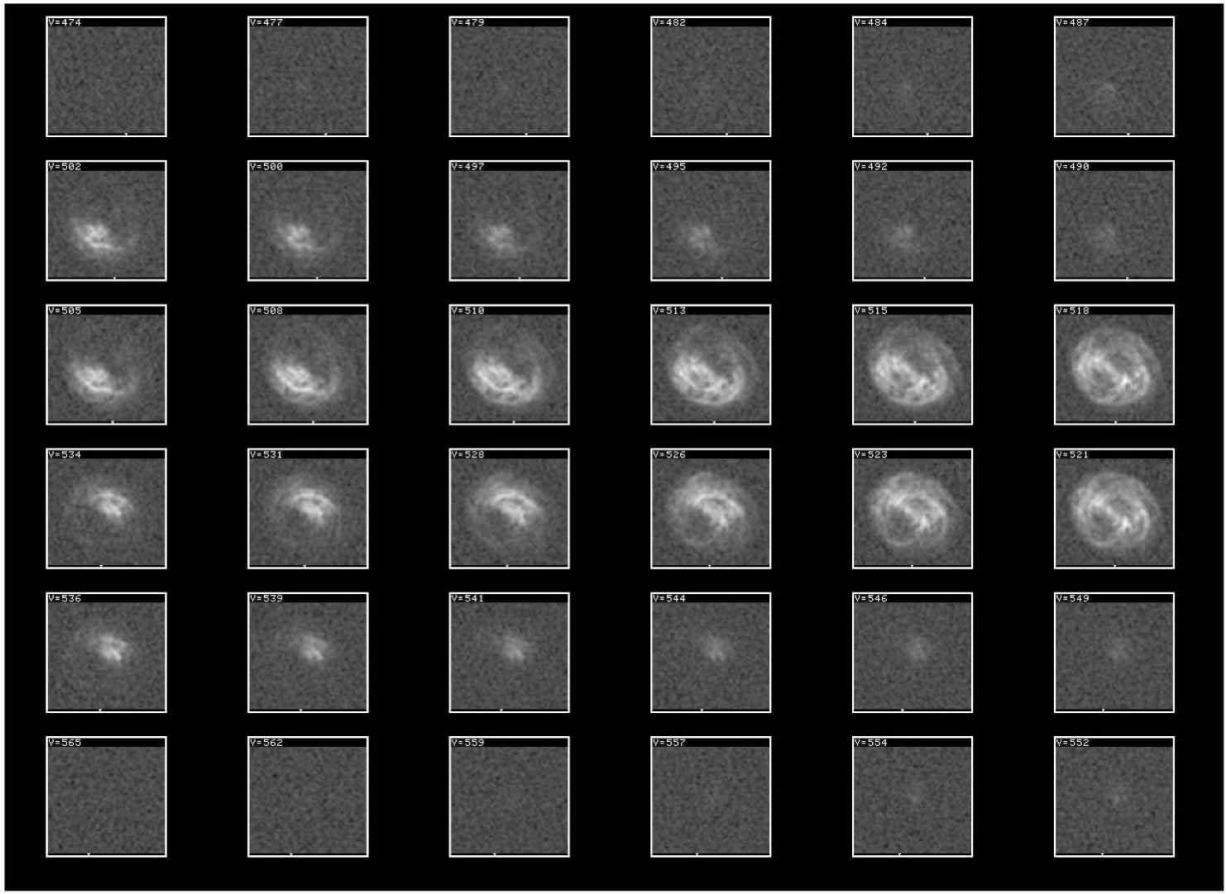}}}
\caption[Sample channel maps for NGC 1058]
{\label{fig:ch_map_1058}Sample channel maps for NGC 1058--Each square image
  represents the \ion{H}{1} 21\,cm line emission within a velocity
  range of {2.58\,km\,s$^{-1}$} where the central velocity of that
  range is given on the upper left corner of each image. The x and y
  axis of every channel map gives the Right Ascension (RA) and
  Declination (Dec) coordinates,and are identical to these in Figure 2. Such sample channel maps can be
  assembled together in the same way the frames of a movie are put
  together to make what is refered to as a data cube.}
\end{figure}
\clearpage

\begin{figure}
\epsscale{.8}
    \plotone{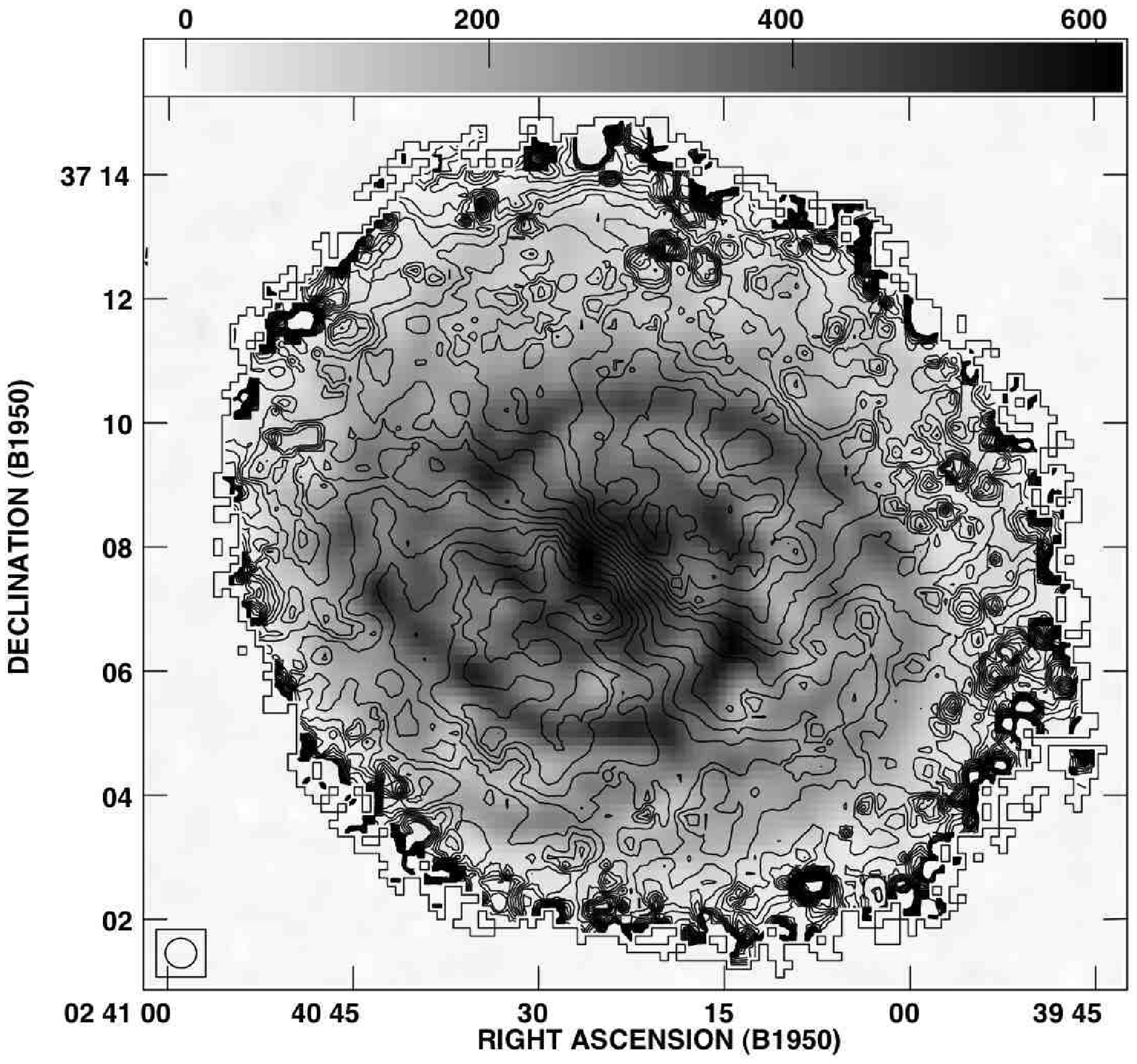}
\epsscale{1}
  \caption{Intensity weighted mean velocity contours atop \ion{H}{1}
    intensity map for NGC 1058
    \label{fig:M1contM0grey}Intensity weighted mean velocity contours atop \ion{H}{1}
    intensity map (grey) for NGC 1058---This figure was made using the
    15\arcsec ~data cube with a sensitivity of 0.5 mJy/beam corresponding to a column density of 
${1.6~\times 10^{18}~\rm{cm}^{-2}}$. The physical resolution of this image is 
    0.7\,kpc and the velocity contours range between 500 and 558 km/sec in 2 km/sec increments. 
     The \ion{H}{1} disk in NGC 1058 extends to a
    diameter of more than 20\,kpc.}
\end{figure}

 \begin{figure}
          \resizebox{\textwidth}{!}{
            \plotone{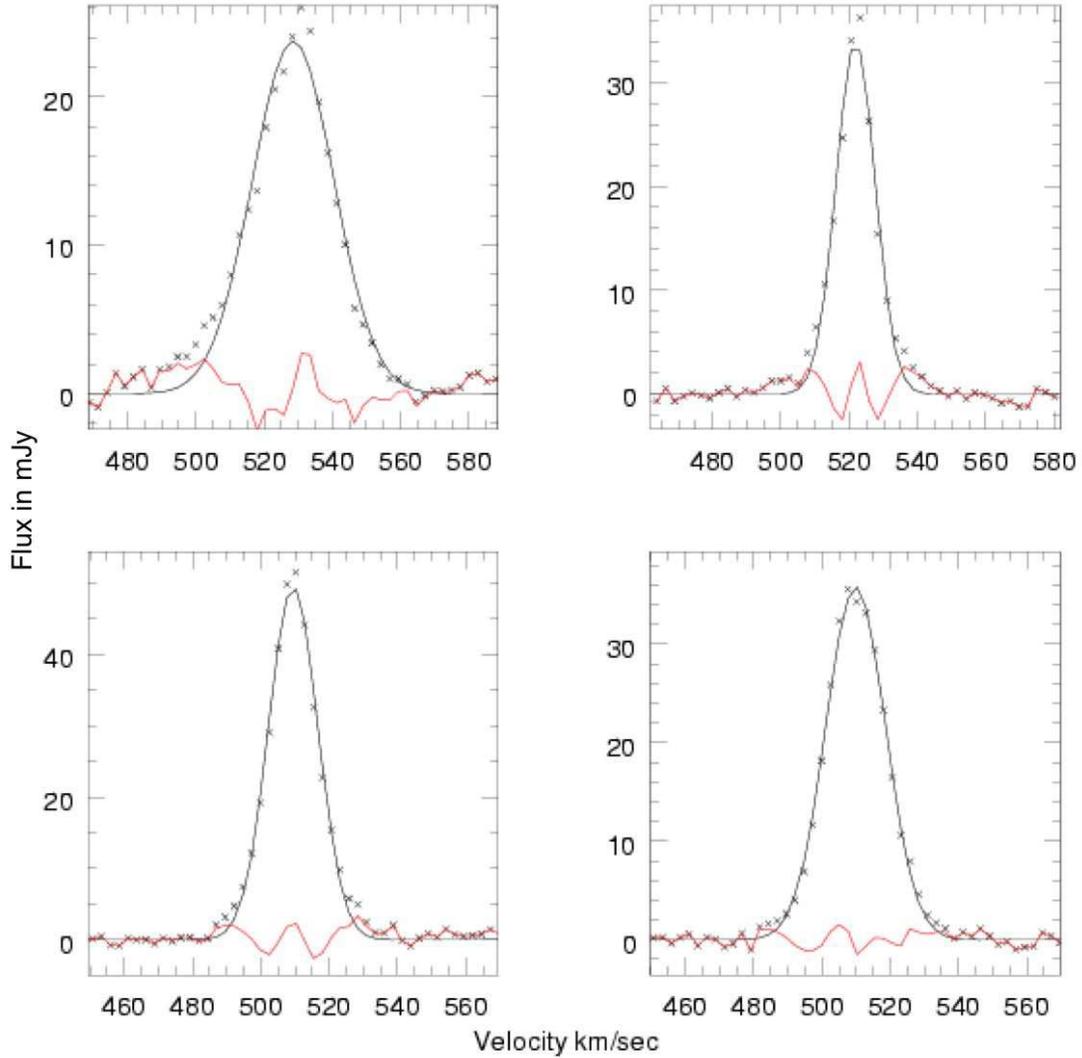}}
\vspace{-3cm}
          \caption[Sample \ion{H}{1} profiles from the 45\arcsec NGC
          1058 data cube]
          {\small \label{fig:prof45_1058}Four sample of
            {\ion{H}{1}} profiles (crosses), the Gaussian fit (solid
            line) and the residual pattern (red) with $\sigma _v$ from
            the Gaussian fit: 12.7\,km/sec (upper left, pixel from a region of high dispersion), 5.95\,km/sec
            (upper right, pixel from a region of low dispersion), 7.6\, km/sec (lower left pixel from an interarm region),
 9.3\,km/sec (lower right, pixel from an arm region).
	    The $x$ axis is in km/sec and the $y$ axis represents {\ion{H}{1}} intensity
in mJy/beam. The shown residuals indicate that a single Gaussian function does not
            adequately describe
            the line shapes. However the width of the Gaussian does
            track the breadth of the {\ion{H}{1}} profile. This figure is
            based on the 45\arcsec ~data cube and the single Gaussian
            fits done for that cube.}
          \end{figure}
          \clearpage
          \begin{figure}
          \resizebox{\textwidth}{!}{
            \rotatebox{-90}{
              \plotone{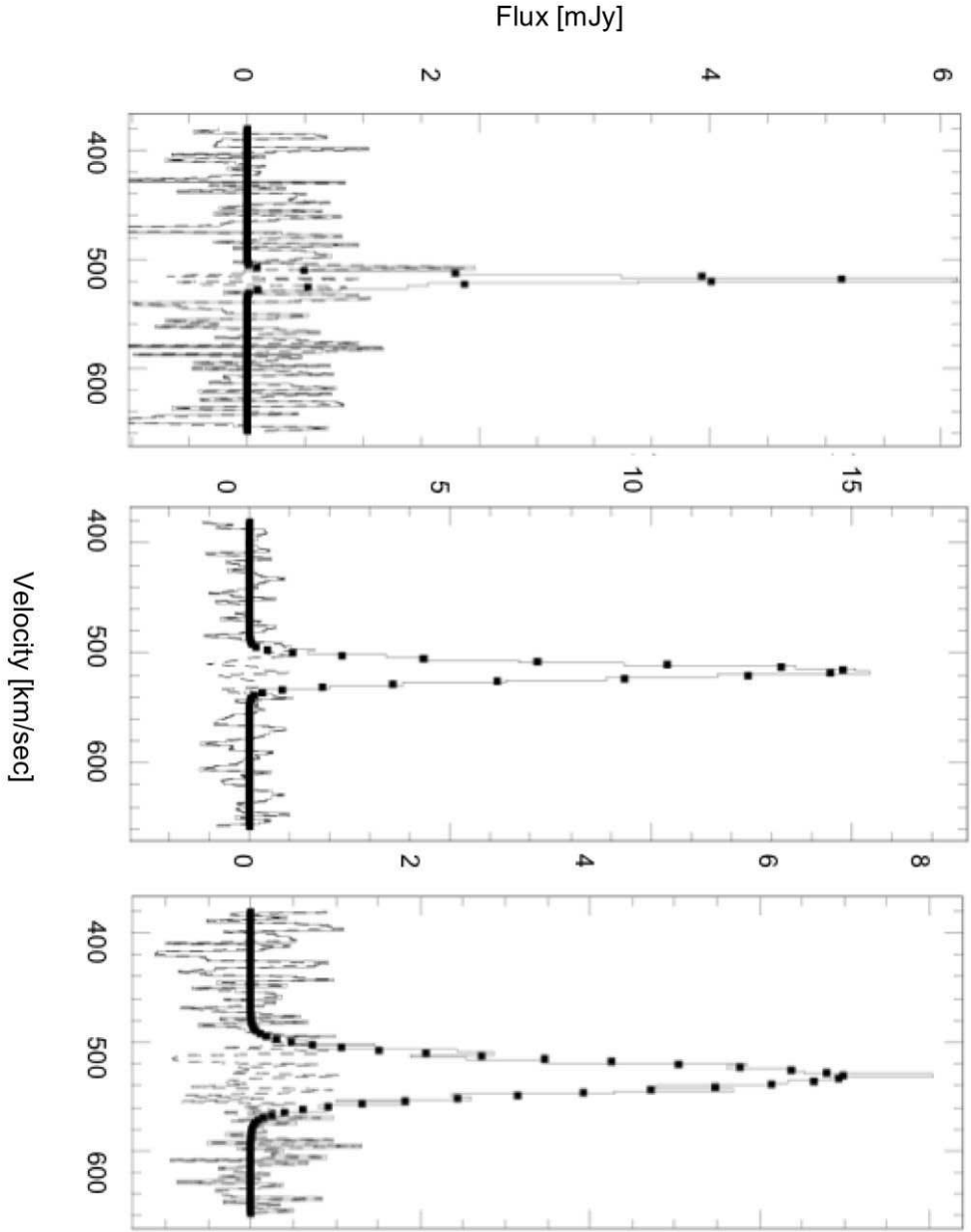}}}
          \caption[Sample \ion{H}{1} profiles from the 30\arcsec NGC
          1058 data cube]
          {\label{fig:prof30_1058} Three samples of {\ion{H}{1}}
            profiles (solid line), the Gaussian fit (squares) and the residual pattern
            (dashed) with $\sigma _v$ from the Gaussian fit: 3.8, 7.6,
            and 13.2  respectively. Figure based on the 30\arcsec ~data 
            cube and the single Gaussian fits done for that cube. The $x$ axis is in km/sec and the $y$ axis represents {\ion{H}{1}} intensity
in mJy/beam.}
        \end{figure}
        \clearpage 

      \begin{figure}
          \resizebox{\textwidth}{!}{
            \plotone{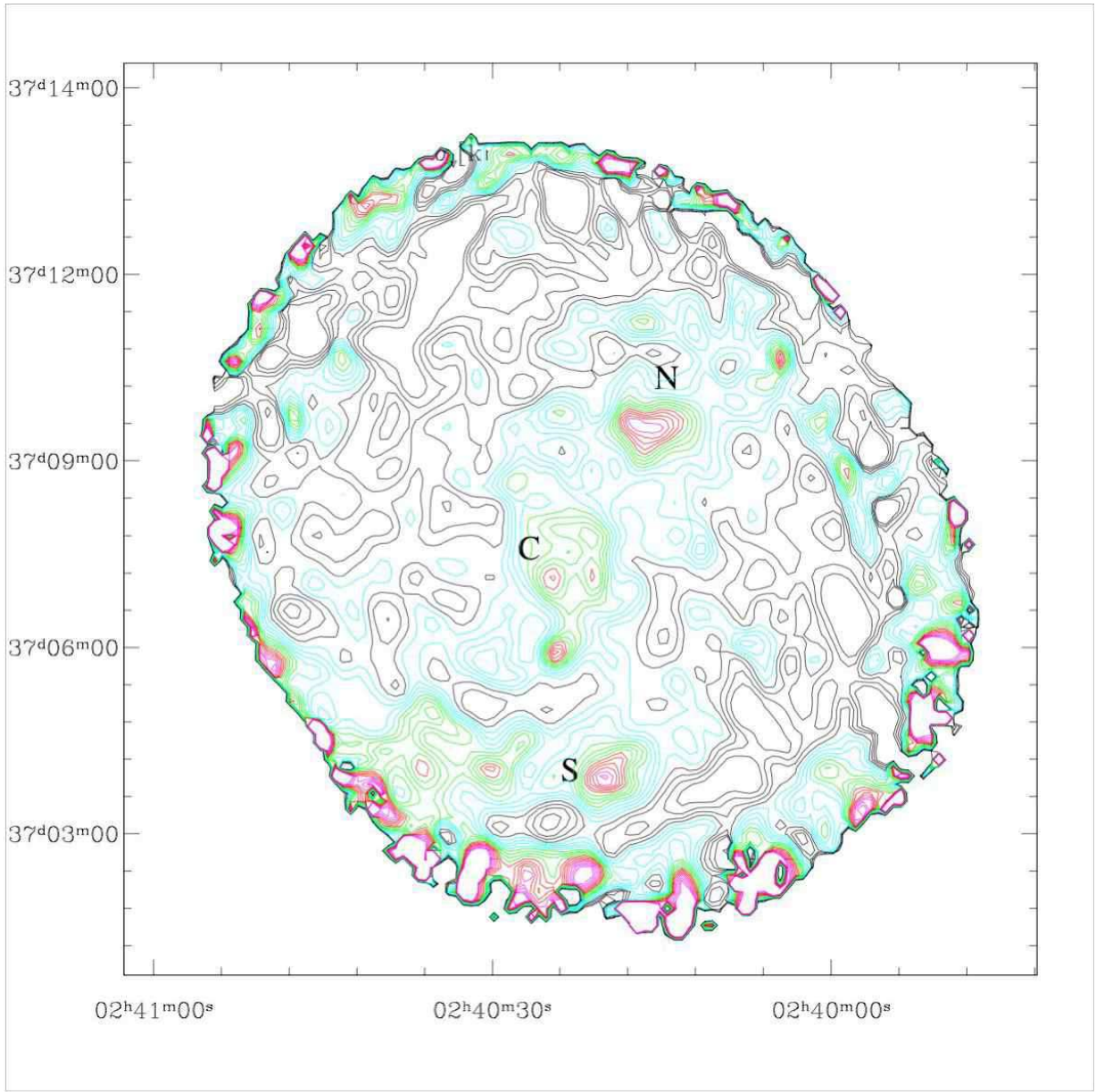}}
          \caption[Distribution of dispersions throughout the
          galaxy]
          {\label{fig:SigCont1058c30} Distribution of dispersions
            throughout NGC~1058; the regions
            of highest dispersion are labeled N, C, and S. The $x$ 
            and $y$ axis are the RA and Dec in B1950 coordinates. This
            figure is based on the results of the single Gaussian fit
            performed on the 30\arcsec ~data. The contours are in km/sec
and start in steps of 0.5 km/sec. Black is used for dispersions between 5.5 to 7, cyan
for 7.5 to 9, green for 9.5 to 11, red for 11.5 to 13, and magenta for 13.5 to 15 km/sec.}
        \end{figure}
        \clearpage

\begin{figure}
          \resizebox{\textwidth}{!}{
                   \plotone{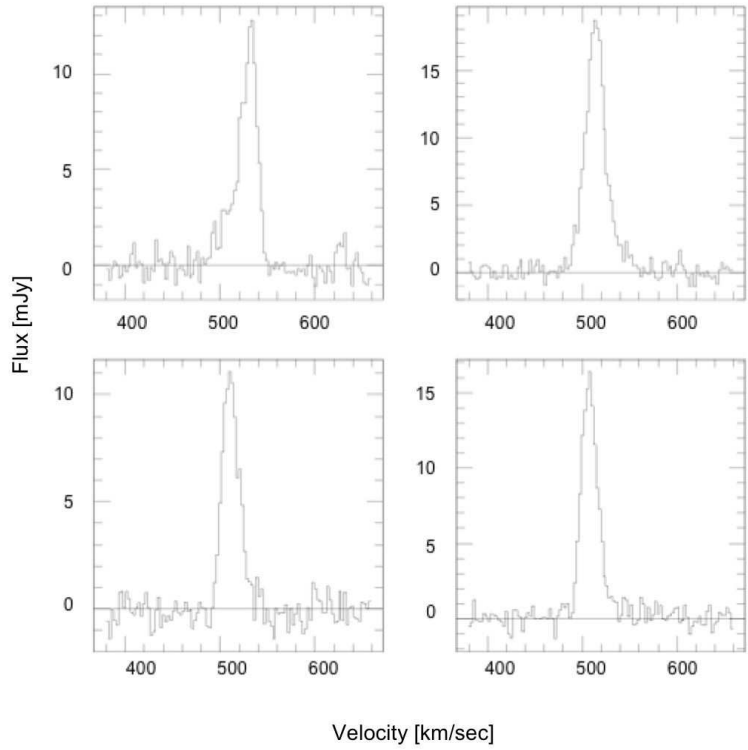}}
\vspace{-2cm}
          \caption[Four sample profiles in the regions with highest asymmetries
 (of order few percent).]{\label{fig:asym}Four sample profiles in the regions with highest asymmetries
 (of order few percent); The $x$ axis of each of the four plots represents velocity and is in units of {km s$^{-1}$} and the $y$ axis represents {\ion{H}{1}} intensity in mJy/beam, where the beam refers to the
point spread function of the observations. The upper left is from region N, upper right from C, lower left from S, and lower right from a region West of S. Regions N,C, and S are shown and labeled in Figure 3.}
        \end{figure}
        \clearpage

        \begin{figure}
          \resizebox{\textwidth}{!}{
          \plotone{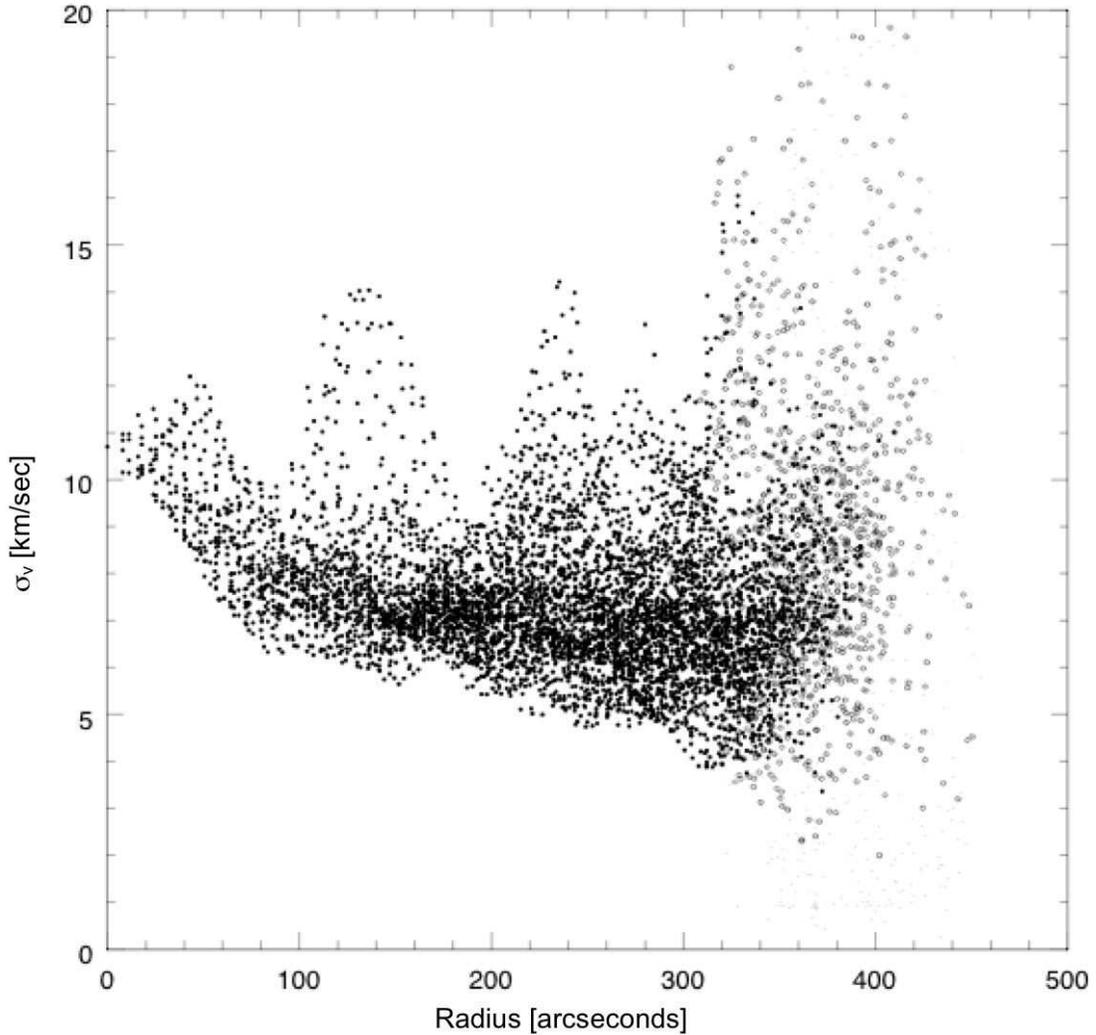}}
\vspace{-3cm}
          \caption[Radial dependence of NGC~1058 $\sigma_v$]
          {\label{fig:SigvsR1058c30} Radial dependece of NGC~1058's 
            $\sigma _v$s---The $x$ axis (in arcseconds) gives the
            radius while the $y$ axis (in km/sec) gives the $\sigma_v$ 
            as derived from single Gaussian least squares fits to the 
            30\arcsec ~data cube. The filled
            circles represent
            points with error bars,less than 12.5\% of $\sigma_v$, the 
            empty circles -- points with error bars between 12.5\% and 25\%
            and the dots-- points with errors greater than
            25\%. Despite a few high $\sigma_v$ regions (N,S in Figure~\ref{fig:SigCont1058c30}, 
the radial falloff is evident.}
        \end{figure}
\clearpage

\begin{figure}
          \resizebox{\textwidth}{!}{
                \plotone{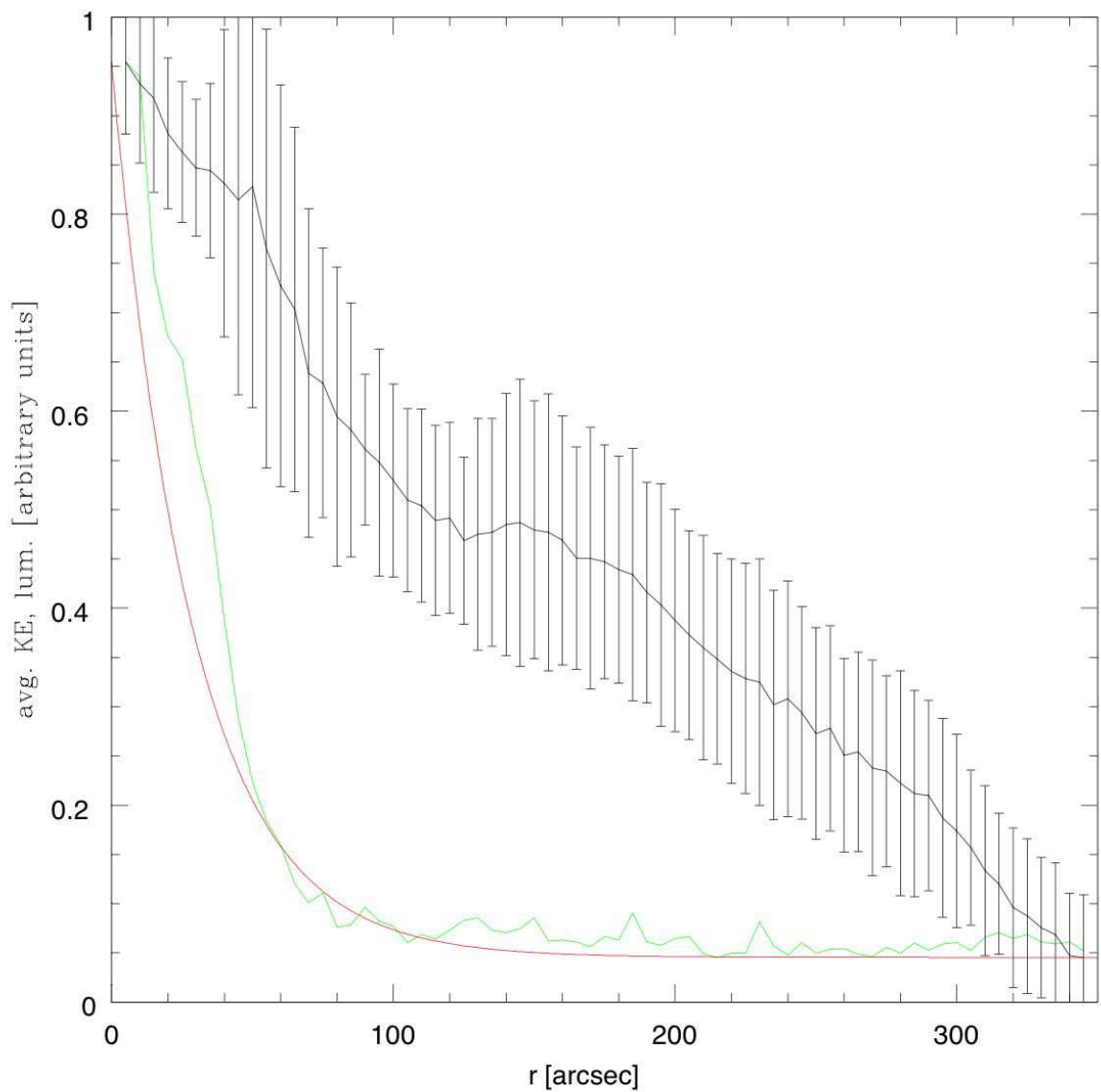}}
          \caption[Qualitative behavior with radius of the kinetic energy in the gas (black) and the
luminosity of NGC 1058 (green).]{\label{fig:ecomp} The energy in the neutral gas
 was roughly approximated as the product of the total \ion{H}{1} intensity and the square of the 
velocity dispersion. Both the green and the black lines represent azimuthal averages of concentric
rings around the center of NGC 1058. The black error bars show the rms in each of these rings. The
red line shows the exponential fit to the stellar data. The energy in the gas falls off 
with radius much slower than the stellar luminosity suggesting that processes other than 
those associated with star input energy are responsible for heating the gas at large radiae.}
        \end{figure}

        \clearpage
       \begin{figure}
          \resizebox{\textwidth}{!}{
	\rotatebox{-90}{
                    \plotone{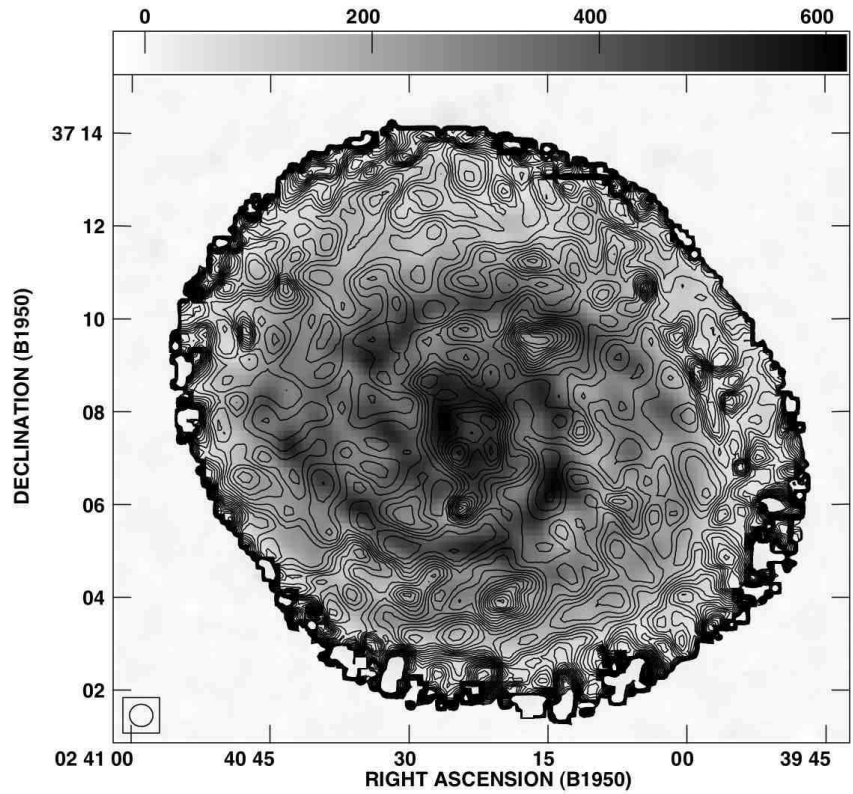}}}
          \caption[$\sigma _v$ contours  atop an {\ion{H}{1}} total
            intensity map]
            {\label{fig:SigonM0c30}$\sigma _v$ contours  atop of an {\ion{H}{1}} total
            intensity map---The contours range from 4 to 14 km/sec in
            steps of 0.5 km/sec and are based on the single Gaussian
            fit to the 30\arcsec , NGC~1058 data cube. Note that regions N and S of high $\sigma _v$
            are located in the inter-arms.}
        \end{figure}
        \clearpage  

 \begin{figure}
          \resizebox{\textwidth}{!}{
	            \plotone{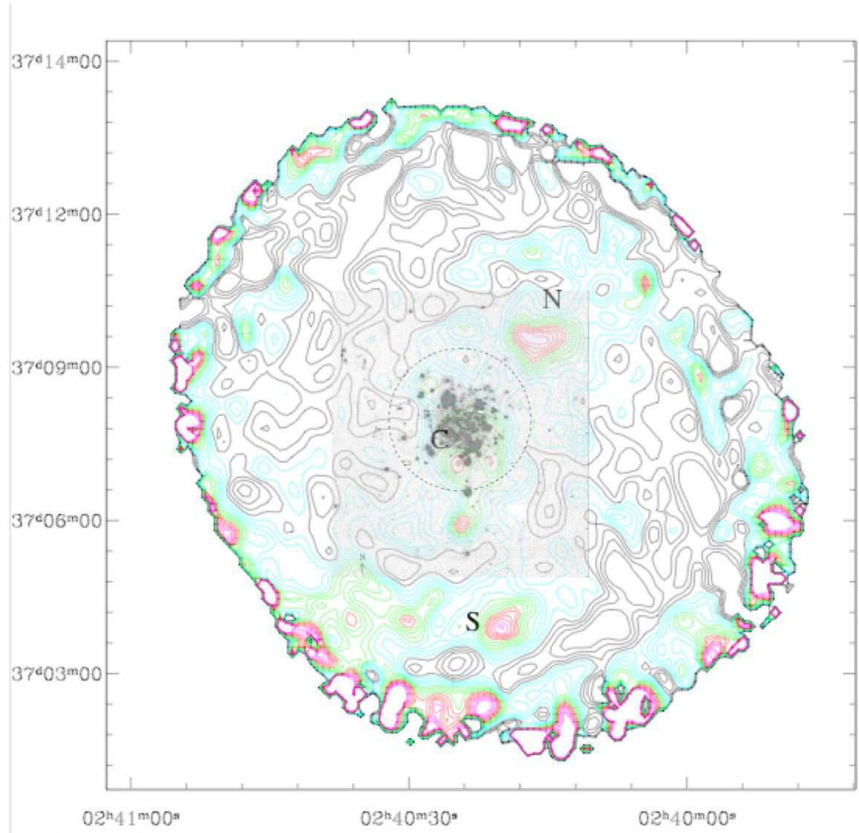}}
          \caption[]
{\label{fig:Halpha_on_Sig} H$_{\alpha}$ greyscale from Ferguson, Gallager, \& Wyse 1998, atop dispersion contours as in Figure 5.}
        \end{figure}
        \clearpage

 \begin{figure}
          \resizebox{\textwidth}{!}{
	            \plotone{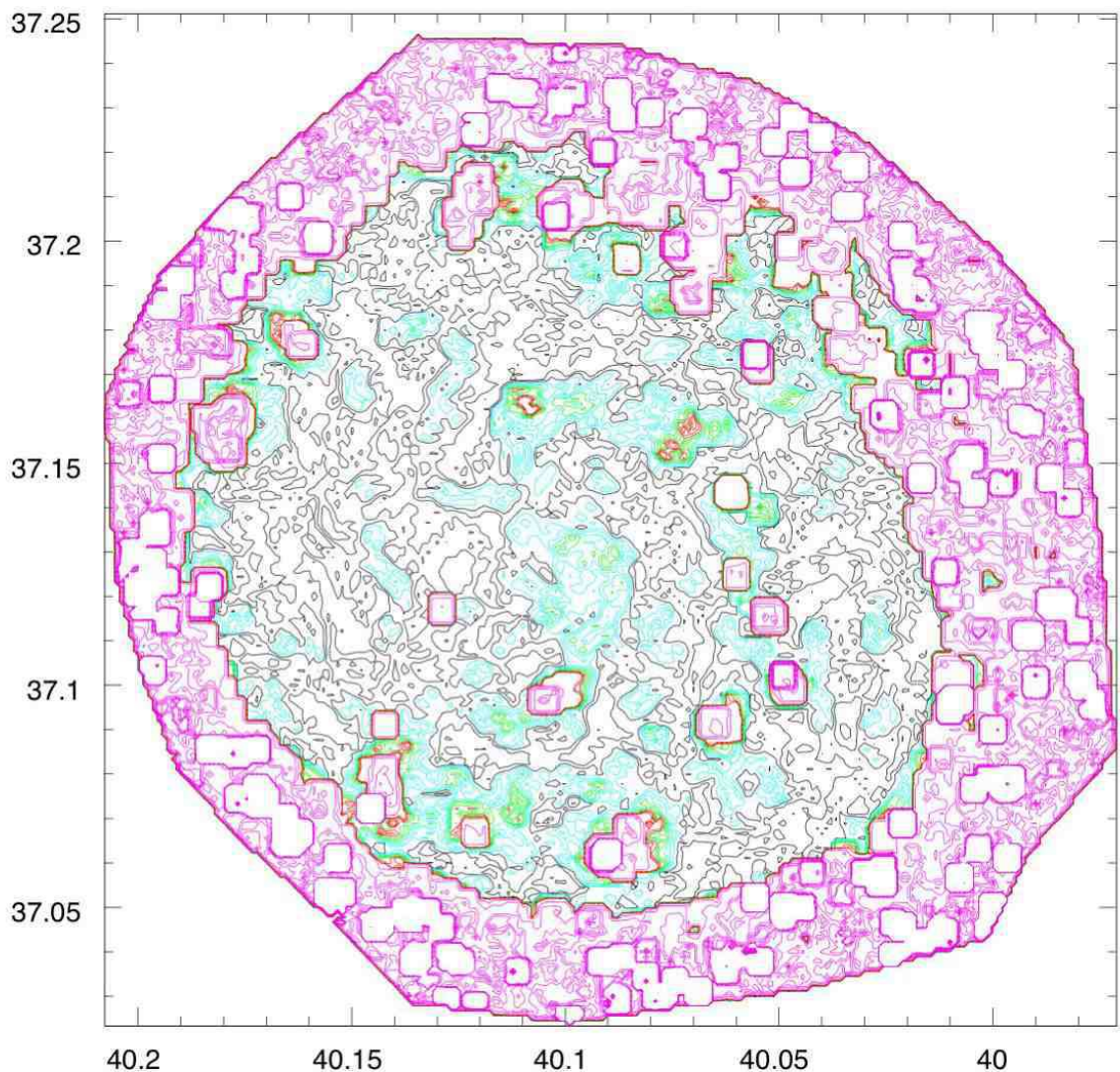}}
          \caption[Map of Maximum Beam Smear at 32\arcsec ~resolution]
          {\label{fig:beam_smear1} Contours of maximum potential beam smearing in km s$^{-1}$. Black is used for dispersions maximum
beam smearing effect of 1, 2, and 3 km/sec, cyan for 4,5,6, green for 7,8,9, red for 10, 11, and 12, 
and magenta for values of 30 km/sec and above. 
 5.5 to 7, cyan for 7.5 to 9, green for 9.5 to 11, red for 11.5 to 13, and magenta for 13.5 to 15 km/sec.
 The magenta contours are regions where the \ion{H}{1} emssion was very
faint or non-existent. As such the Gaussian fitting routine employed produced spurious results. The
square shape of some of the contours is an artifact of the method employed in determining the maximum 
beam smearing. This figure is based on the 15\arcsec ~data cube. Note that the 
highest velocity gradients are found in regions N and S and south-west of C. Regions N, S, and C are
shown and labeled in Figure 5.}
        \end{figure}
        \clearpage

 \begin{figure}
          \resizebox{\textwidth}{!}{
            \plotone{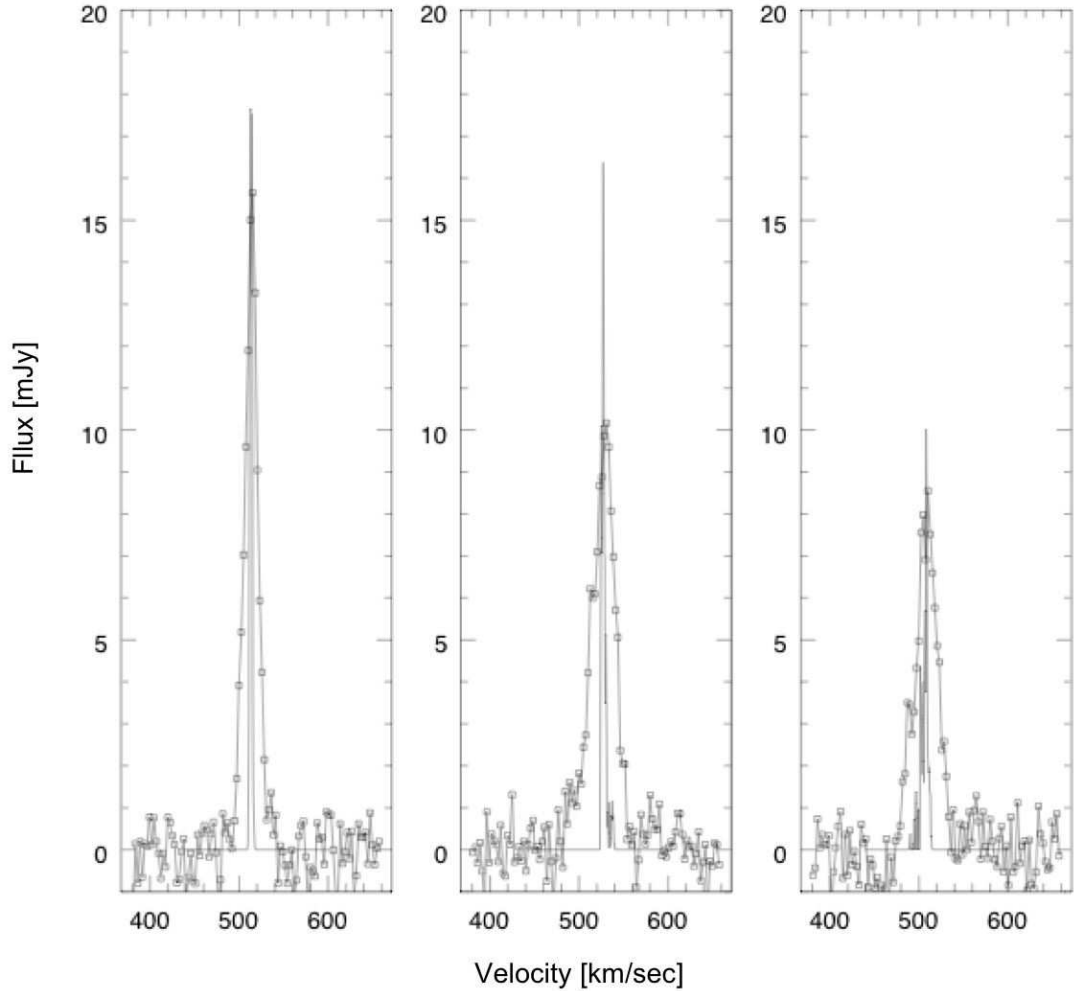}}
\vspace{-2cm}          
\caption[]
          {\label{fig:beam_smear2} Effect of beam smearing on $\sigma _{v}$ measurements at 30\arcsec 
~resolution. The $x$ axis in in km/sec and the $y$ axis is in mJy/beam. The connected squares represent the observed profile. The narrow profiles
were obtained by modeling the velocity profiles associated with an infinitely
cold disk and then convolving that model with a 30\arcsec ~beam, and running the Gaussian least squares
fitting routine on the convolved cube.}
        \end{figure}
        \clearpage

\begin{figure}
\resizebox{\textwidth}{!}{	
\plotone{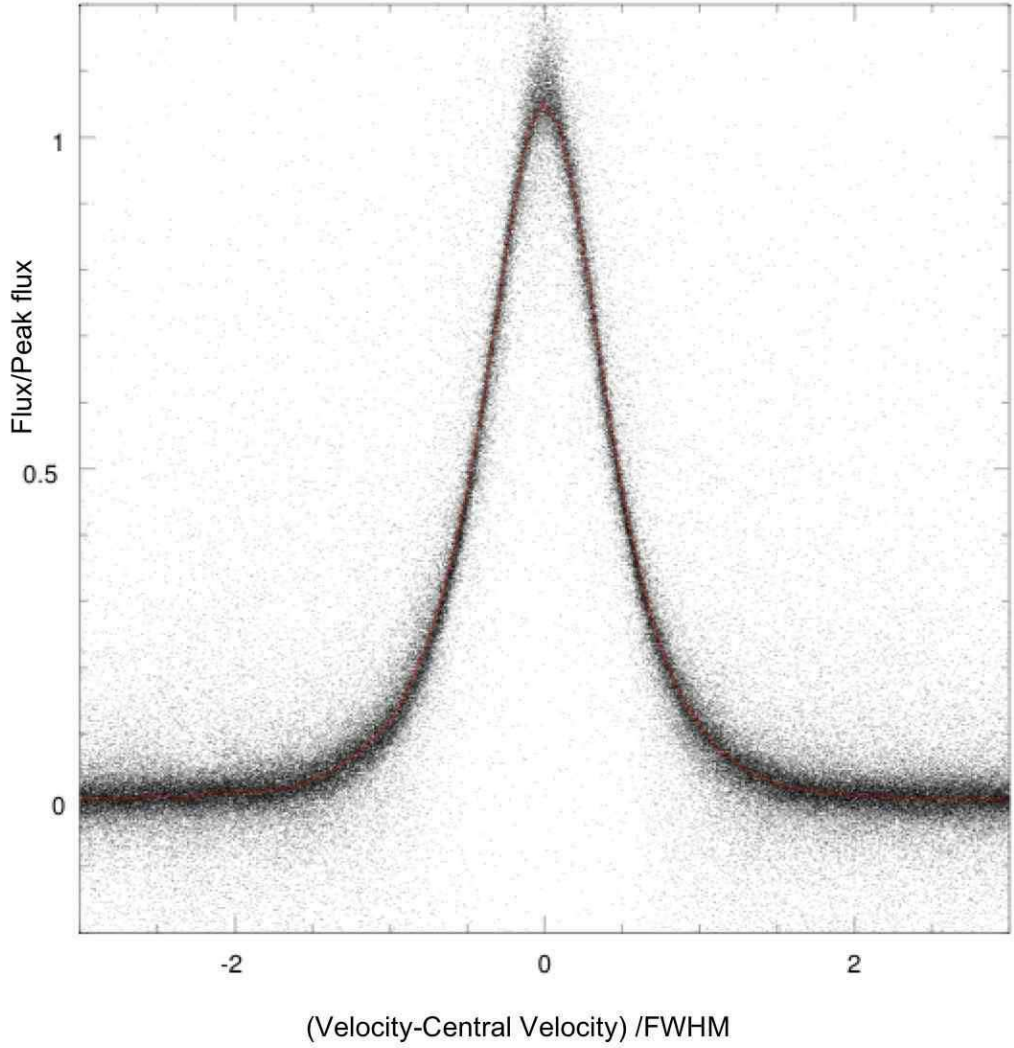}}
\caption[Median profile atop all \ion{H}{1} profiles from NGC~1058
45\arcsec data cube]
{\label{fig:UPfid45}Median Profile (red) atop all \ion{H}{1} profiles from
  NGC~1058 45\arcsec ~data cube. The $x$ axis is flux/peak flux and the $y$ axis is
velocity minus the central velocity and divided by the FWHM.}
\end{figure}

\begin{figure}
  \resizebox{\textwidth}{!}{
    {\plottwo{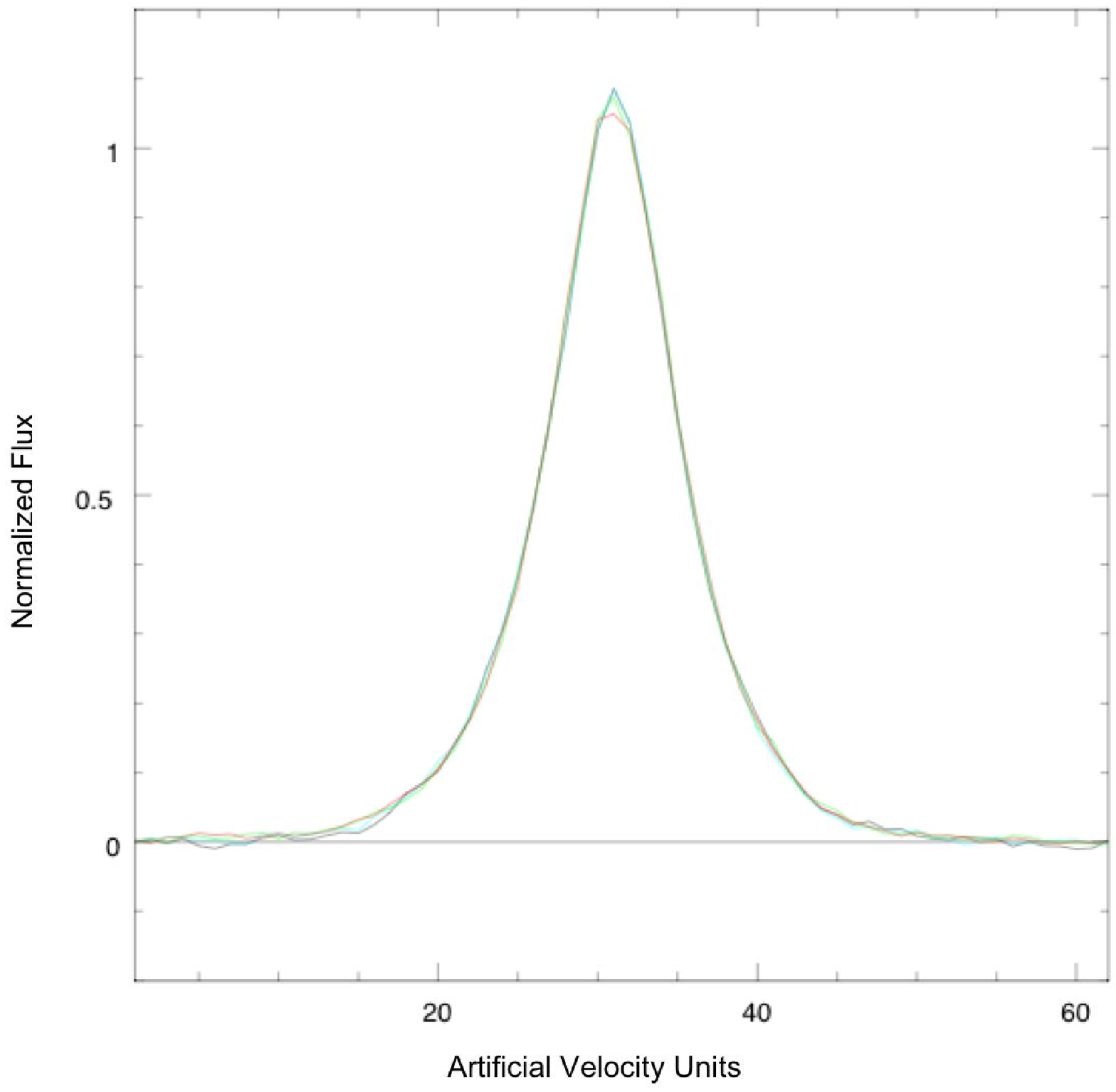}{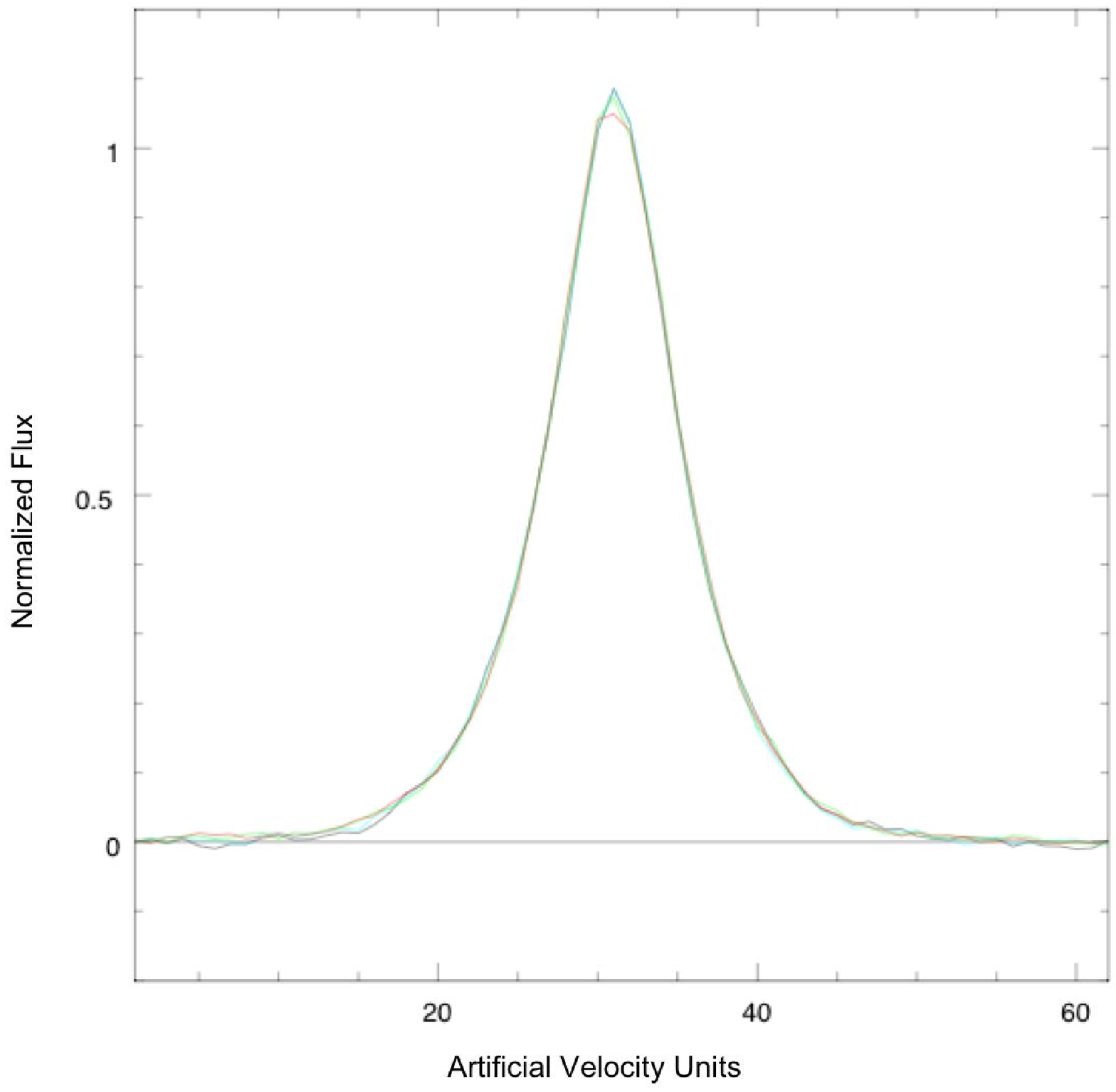}}}
  \caption[]
    {\label{fig:up45test} Median profiles derived for certain FWHM
      ranges (left pannel);and for peak flux ranges (right pannel) from the 45\arcsec 
~NGC~1058 data set. The $x$ axis is not in [km/sec] but represents
 the grid (bin number) on which the profiles were set. Please refer to text for a detailed
explanation. The $y$ axis is the flux divided by peak flux. In the left pannel black is used for
median profiles with widths (FWHM) between 14 to 18 km/sec, cyan for 18 to 22 km/sec, green
22 and red for 26 to 30 km/sec. In the right pannel a solid line is used for profiles
with peak fluxes between 10 and 25 mJy/beam, a dotted line is used for profiles
with peaks between 25 and 40 mJy/beam, short dash for 40 to 55, and long dash for
55 to 70 mJy/beam. }
\end{figure}
\clearpage

\begin{figure}
  \resizebox{\textwidth}{!}{
    \plotone{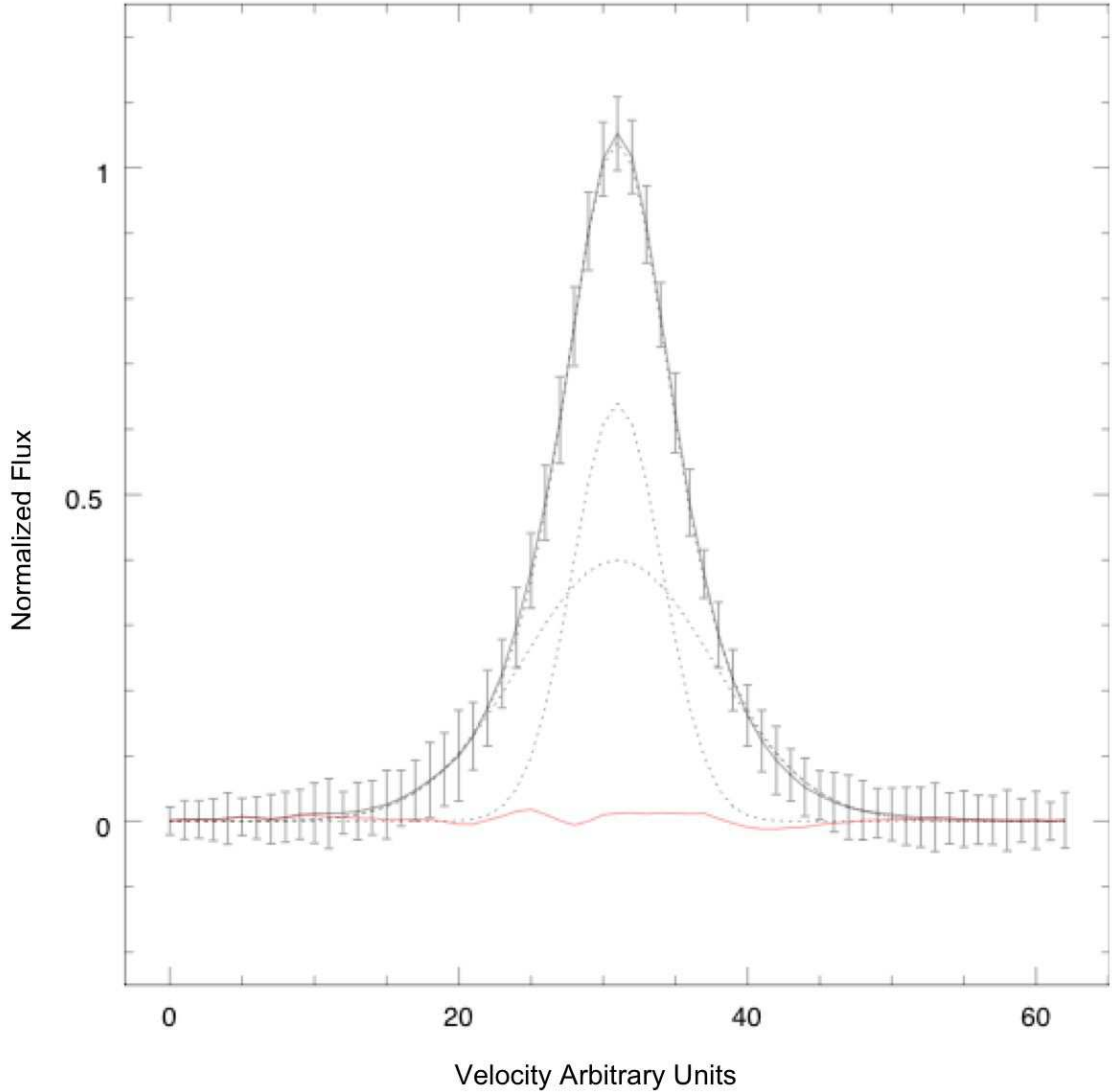}}
\vspace{-2cm}
  \caption[Double Gaussian Fit to the Universal Profile in NGC~1058]
  {\label{fig:UP2G}Double Gaussian Fit to the Universal Profile --- The $x$
    axis is in bins as described in the text and the $y$ axis is the
    flux normalized to the peak intensity. The two
    Gaussian components used to fit the Median profile and their sum
    are shown in dotted lines. The residuals are shown in red. Error
    bars based on the rms in each bin are
    also given. The ratio between the areas of the broad and narrow
    components is 1.35 while that between their FWHM is 2.09.}
\end{figure}
\clearpage

\begin{figure}
\resizebox{\textwidth}{!}{
  \plotone{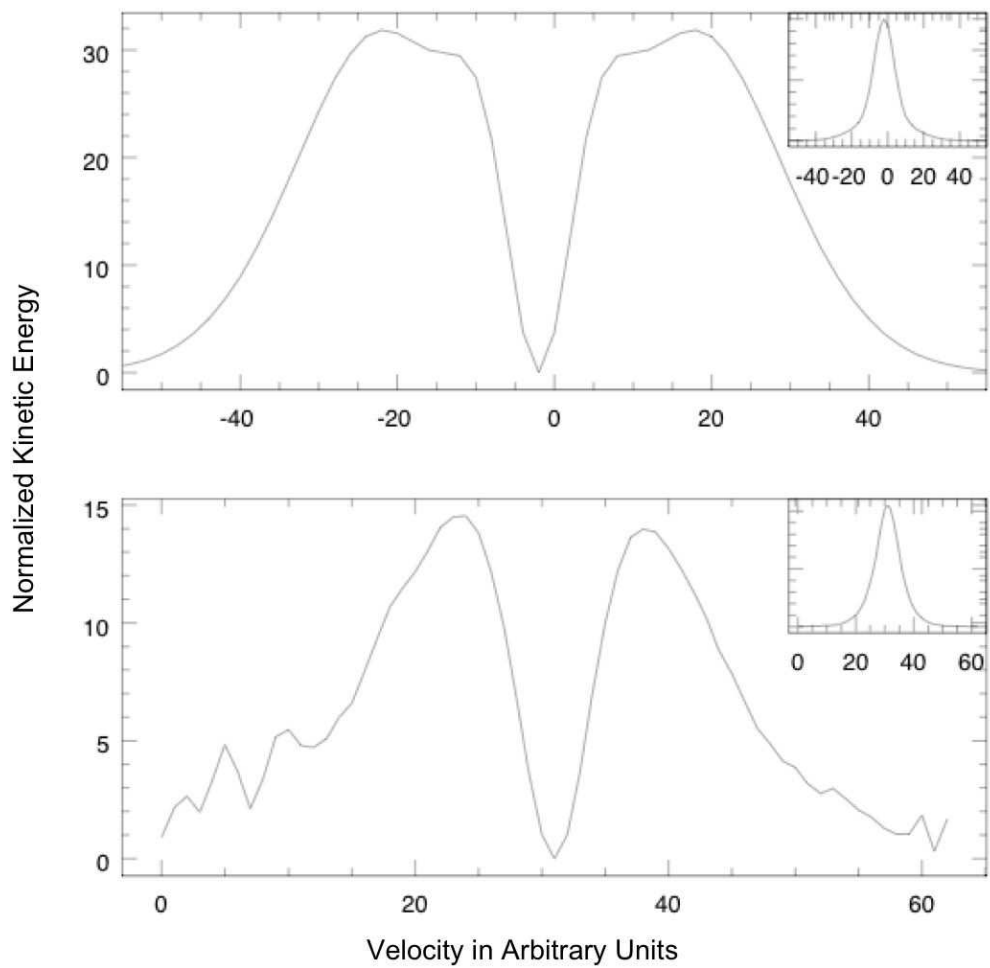}}
\vspace{-5cm}
\caption[Normalized Kinetic Energy distributions for the Milky Way and
  NGC 1058 and corresponding \ion{H}{1} profiles]
  {\label{fig:UPenergy} Normalized Kinetic Energy distributions for
 the Milky Way (top) and NGC 1058 (bottom) and corresponding \ion{H}{1} profiles. 
The top figure was obtained from a double Gaussian decomposition of the
 North Galactic Pole \ion{H}{1} emission, from Kulkarni \& Fich (1985).
  For the top figure, the $x$ axis represents
  velocity in units of {km\,s$^{-1}$ and the $y$ axis is the normalized energy 
  in units of {Kelvin km$^2$ s$^{-2}$}}. The inset upper right figure
  shows the \ion{H}{1} profile from which the normalized energy curve
  for the Milky Way's North Galactic Pole emission was estimated.
  The bottom figure shows the qualitative behaviour of the kinetic
  energy distribution with velocity in NGC 1058. Here  
  the $x$ axis represents a bin number (ref.to text) while the
  $y$ axis represents the qualitative behaviour of the kinetic energy
  distribution in NGC 1058. This figure suggests that the kinetic energy
  in the Galactic North Galactic Pole emission is
  more evenly distributed in velocity than that in NGC 1058. 
  }
\end{figure}
\clearpage
\end{document}